\documentclass{vgtc}                          

\graphicspath{{figures/}{pictures/}{images/}{./}} 

\usepackage{times}                     

\usepackage{tabu}                      
\usepackage{booktabs}                  
\usepackage{lipsum}                    
\usepackage{mwe}                       

\usepackage{mathptmx}                  

\newcommand{\langlasso}{\textit{LangLasso~}}

\onlineid{0}

\vgtccategory{Research}

\vgtcinsertpkg

\title{LangLasso: Interactive Cluster Descriptions through LLM Explanation}

\author{
Raphael Buchmüller\thanks{e-mail: raphael.buchmueller@uni.kn}\\ %
\scriptsize University of Konstanz %
\and Dennis Collaris\thanks{e-mail: d.a.c.collaris@uu.nl}\\ %
\scriptsize Utrecht University %
\and Linhao Meng\thanks{e-mail: l.meng1@tue.nl}\\ %
\scriptsize Eindhoven University of Technology %
\and Angelos Chatzimparmpas\thanks{e-mail: a.chatzimparmpas@uu.nl}\\ %
\scriptsize Utrecht University %
}

\teaser{
\vspace{-4mm}
  \centering
  \includegraphics[width=\linewidth]{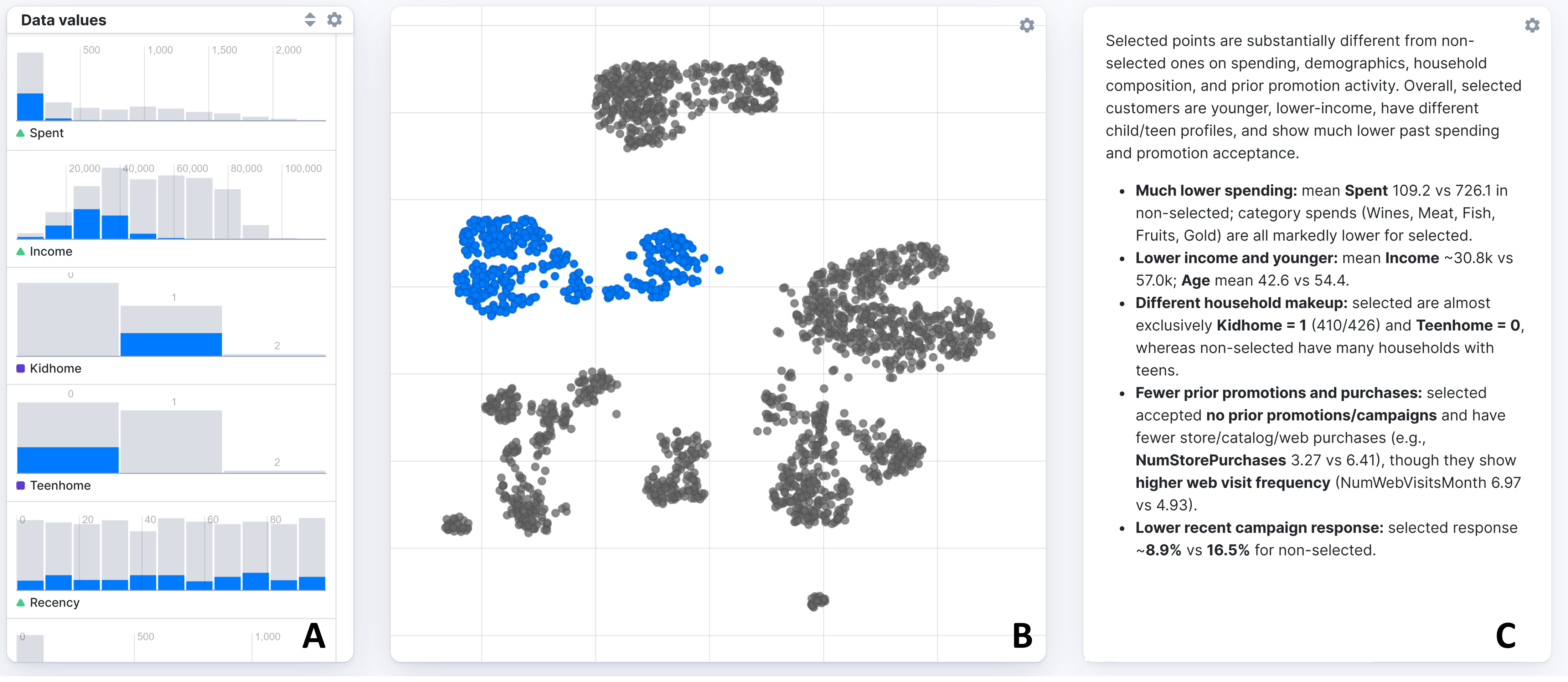}
  \vspace{-7.5mm}
\caption{\langlasso combines visual inspection with language-based cluster explanations. (A) Attribute views enable manual exploration of feature distributions but often require scrolling through many dimensions. (B) A 2D UMAP projection reveals high-dimensional patterns and supports lasso selection of clusters of interest. (C) For the selected cluster, \langlasso generates natural-language summaries highlighting key distinctions from the rest of the data, streamlining interpretation compared to (A).}
  \label{fig:teaser}
}

\abstract{
Dimensionality reduction is a powerful technique for revealing structure and potential clusters in data. However, as the axes are complex, non-linear combinations of features, they often lack semantic interpretability. Existing visual analytics (VA) methods support cluster interpretation through feature comparison and interactive exploration, but they require technical expertise and intense human effort. We present \textit{LangLasso}, a novel method that complements VA approaches through interactive, natural language descriptions of clusters using large language models (LLMs). It produces human-readable descriptions that make cluster interpretation accessible to non-experts and allow integration of external contextual knowledge beyond the dataset. We systematically evaluate the reliability of these explanations and demonstrate that \langlasso provides an effective first step for engaging broader audiences in cluster interpretation. The tool is available at \href{https://langlasso.vercel.app/}{langlasso.vercel.app}.
}


\begin{document}
\maketitle

\section{Introduction}


A staple of many visual analytics (VA) approaches to understand multidimensional data is to use dimensionality reduction techniques (e.g., t-SNE, UMAP) to obtain a lower-dimensional representation~\cite{ren2025embedding, ren2025scalable}. This transformation not only helps to visualize the complex data but can also reveal the underlying structure and potential clustering in the data.
%
Despite their effectiveness, these complex techniques yield non-linear transformations that obscure the semantic meaning of the original features; while structure and clusters are revealed, the axes are difficult to interpret, and it remains challenging to figure out what constitutes a cluster. 

\begin{figure*}[ht]
  \centering
  \includegraphics[width=0.87\textwidth]{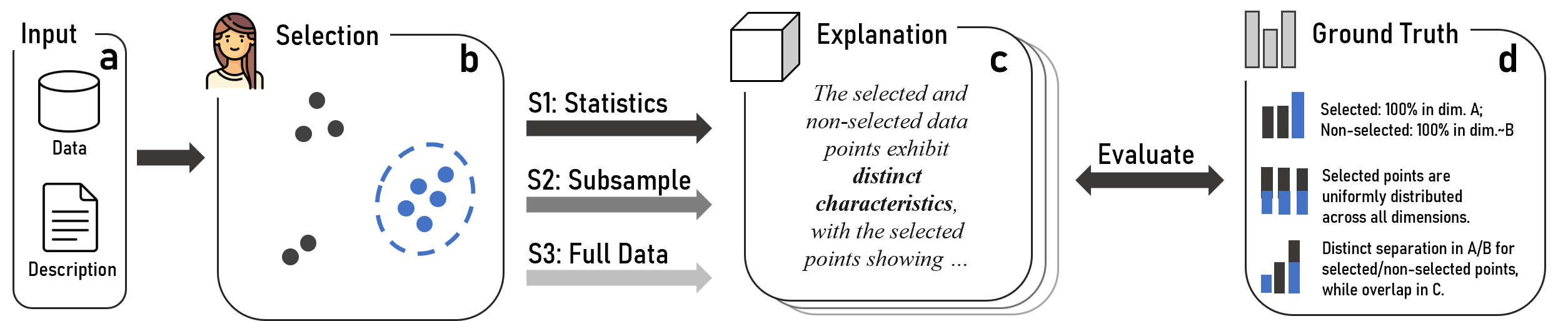}
    \vspace{-3mm}
\caption{Pipeline of \textit{LangLasso}. (a) We upload a dataset and a file with per-feature descriptions. (b) We select a cluster in the projection and test three strategies (S1--S3). (c) \langlasso generates a natural-language explanation of differences between selected and non-selected points. (d) We evaluate the explanations using feature distributions, prior human findings, and/or statistics computed from the data.}
  \label{fig:pipeline}\vspace{-3mm}
\end{figure*}

VA approaches can help understand data clusters by: conducting in-depth analysis; interactively exploring the clusters through brushing and linking multiple views; and comparing the feature distributions in the data and the various clusters to show which are most responsible for cluster formation. 
%
This approach has been successful in \cite{Fujiwara2020Supporting} but requires considerable time and technical expertise, limiting accessibility for non-experts outside the analytical domain.

We present \textit{LangLasso}, an interactive approach that uses LLMs to generate natural language descriptions of clusters (Fig.~\ref{fig:teaser}). By offering human-readable explanations, even non-technical (or less visually-literate) users have a means to interpret clusters. \langlasso is an entry point for these users to engage with VA approaches, explore the clusters further, and test their hypotheses. LLMs can also incorporate external contextual information beyond the original dataset to construct more intuitive explanations~\cite{Peters2024Large}. For example, an LLM can infer social media users' psychological dispositions from their posts by using linguistic patterns and world knowledge, even without explicit training on those traits~\cite{Peters2024Large}. 

Although natural language is often easier to understand, it is limited in the nuance and detail it can convey. It is also subject to potential hallucinations since the interpretation is based on an LLM. To address these issues, we conduct a systematic evaluation to report on the reliability of our approach. We also advocate for using our approach in conjunction with (or to enhance) existing approaches, rather than relying on LLM output in isolation. Our evaluation shows that the statistics strategy is the most reliable approach, providing precise and reproducible cluster descriptions (by comparing against the statistics we computed for the LLM), while the remaining strategies reduce accuracy and interpretability.


\section{Related Work}
\noindent\textbf{Reliability Considerations --} In VA, it is common to project high-dimensional data into a 2D space and display the results as scatterplots, where spatial proximity suggests similarity. Visually emergent clusters often represent meaningful patterns that users identify and explore during interaction with these projections. Explaining such clusters involves uncovering the key features or patterns that distinguish one region of the projection from another. A common approach is to determine which input dimensions contribute most to these visual separations. Traditional methods typically rely on statistical comparisons of feature values or distributions within and across selected regions~\cite{10.2312:eurova.20151100,TIAN202193}. Other works utilize projection enrichments~\cite{10.2312:eurova.20241111} or representative instances, such as centroids or exemplars~\cite{doi:10.1137/1.9781611978032.6}, to summarize clusters and highlight distinguishing features. Additionally, several interactive visualization tools have been developed to support the analysis and interpretation of projection results~\cite{7192695,9064929}. These tools often present feature distributions for selected regions and enable users to visually compare them. However, such visual comparisons impose high perceptual load—especially with many dimensions or weak cues. Complementing this, \emph{Explain-and-Test} generates regional explanations and validates them via a local classifier and re-projection, linking why a cluster forms to whether it holds~\cite{explain_raval}. LLM-based cluster naming has likewise been benchmarked (e.g., GPT-3.5-turbo vs. human labels), showing feasible but limited automation~\cite{preiss_2024}.

\noindent\textbf{Natural Language for Visualization --} The emerging research area, natural language for visualization, focuses on the integration of natural language processing techniques, including LLMs, with visualization systems, enabling more intuitive and accessible data exploration~\cite{2023NL4VISsurvey}. With the emergence of generative models, one promising direction is to generate visualizations from textual input~\cite{2023Chat2VIS, 2024testNL4vis}. Natural language is also being used to support the interpretation of visual content, helping users understand complex visual encodings. Hong et al.~\cite{2025NL4VisLiteracy} investigate how LLMs can explain static visualizations in natural language, generating descriptions and insights that improve visualization literacy and accessibility for broader audiences. Beyond visualization generation and understanding, LLMs are being applied to assist with visualization-specific tasks such as editing or querying structured visual formats~\cite{2024LLM4SVGVis}, summarizing texts when exploring document-based data~\cite{2025PromptLenses}, and generating chart descriptions~\cite{MatplotAlt}. LLMs are also fine-tuned for table summarization tasks in AI contexts~\cite{2024XingTableLLMSpecialist}. Mo et al.~\cite{2025TableNarrator} present TableNarrator, a tool that uses LLMs to bridge the gap between numeric tables and narrative explanation.  In this work, however, we focus on using natural language to summarize data characteristics of tabular data for supporting data interpretation when interacting with 2D visual embeddings. 


\section{Method}
\noindent\textbf{Design Rationale --} \langlasso is conceived as an additional component to established VA workflows. It targets novice users who are accustomed to static visualizations and may lack expertise in visualization or data analysis, such as business professionals and decision makers. The goal is to provide fast, low-effort interpretation of clusters in dimensionality reduction plots. It extends brushing and linking techniques with concise, language-based summaries that are easier to interpret, thereby lowering the entry barrier to cluster analysis while preserving compatibility with existing VA methods \cite{Collaris2023StrategyAtlas}.


\noindent\textbf{Method Description --} 
We begin the evaluation by providing the LLM with the dataset, together with a description of its domain and dimensions, to supply the necessary context for the reasoning and interpretation tasks (Fig.~\ref{fig:pipeline}a). A 2D projection is then computed to support selecting a cluster of interest (b). We use UMAP (nNeighbors=50, minDist=0.6, spread=2) with deterministic seeding and progressive updates, which yields compact, well-separated clusters for interactive selection and interpretation~\cite{8851280}. 
For each selection, \langlasso assembles evidence for prompting. We compare three strategies that vary the selection-specific information passed to GPT-5-mini, a faster, more cost-efficient version of GPT-5, allowing us to verify our findings relative to the maximum capabilities of current LLMs. The dataset description, strategy-specific input data, and corresponding task prompt (specified in Sec.~\ref{sec:strategies}) are composed into a structured query. The LLM returns a live textual description of the user selection (Fig.~\ref{fig:pipeline}c and Fig.~\ref{fig:teaser}C). Finally, we qualitatively evaluate the LLM outputs against ground-truth statistics and single-feature distributions (Fig.~\ref{fig:pipeline}d and Fig.~\ref{fig:teaser}A).


\noindent\textbf{Strategies --} \label{sec:strategies} We consider three strategies, each providing different information about the selected and non-selected sets to the LLM. The statistics strategy (\textit{S1}) relies on precomputed summary statistics: numerical features include measures such as the minimum, maximum, mean, standard deviation, and Kolmogorov–Smirnov (KS) statistic, while categorical features report category counts and a KS statistic. The subsample strategy (\textit{S2}) uses raw tabular data, sampling 20\% of the selected and non-selected sets. The full data strategy (\textit{S3}) provides the complete data for both selected and non-selected sets.
For each strategy, we construct a task prompt that begins with a fixed instruction: ``I want you to act as a data analyst...''. This is followed by a description of the data provided for the selected strategy, a task description, and the expected response format (a short summary plus 3--5 bullet points, under 200 words, with key terms in bold). \textit{S1} focuses on comparing aggregated statistics, offering efficiency and precision but with limited granularity. \textit{S2} and \textit{S3} provide raw feature values, requiring the LLM to perform the underlying analysis directly and generate the corresponding explanations.







\begin{table}[ht]
\centering
\caption{Datasets with meaningful attributes used in our experiments.}
\vspace{-1mm}

\scalebox{0.85}{ 
\begin{minipage}{\linewidth}
\renewcommand{\arraystretch}{0.9}
\setlength{\tabcolsep}{4pt}
\begin{tabular}{lccc}
\hline
\textbf{Dataset Name} & \textbf{\# Samples} & \textbf{\# Attr.} & \textbf{\# Labels} \\
\hline
Palmer Penguins~\cite{Gorman2014Ecological} & 333 & 6 & 3 classes \\
Bank Marketing~\cite{Moro2014A} & 11,162 & 18 & 2 classes \\
Food Nutrition~\cite{usda_fooddata_central} & 7,499 & 12 & 7 clusters \\
Customer Analysis~\cite{customer_analysis_data} & 2,212 & 31 & 4 clusters \\
\hline
\end{tabular}
\end{minipage}
}
\label{tab:all_meaningful_datasets}
\end{table}

\noindent\textbf{Datasets --}
To evaluate our approach, we selected four datasets from different domains, with varying numbers of samples and attributes (cf. Table~\ref{tab:all_meaningful_datasets}). The first two datasets are classification problems with known class labels, enabling us to verify our proposed strategies (see Sec.~\ref{sec:strategies}) and examine specific clusters by analyzing the distributions of important features. The last two datasets are clustering problems without predefined labels, where our analysis emphasizes comparisons with narrative findings derived from prior works. This selection ensures diversity not only in dataset size and attribute dimensionality, but also in data structure and analytical perspective.


\section{Results}

\begin{figure}[ht]
  \centering
  \includegraphics[width=0.5\columnwidth]{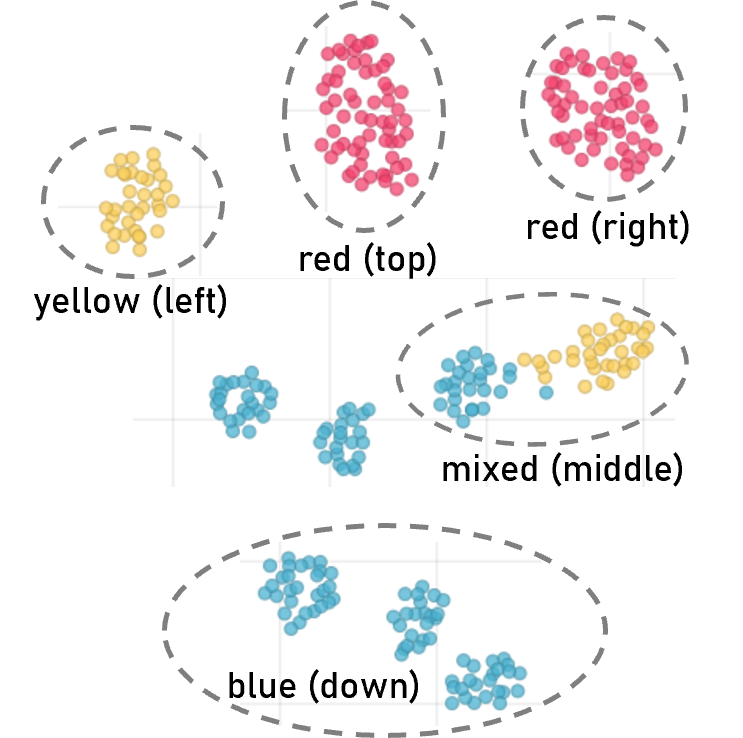}
    \vspace{-3mm}
  \caption{Penguins' projection with three classes but many clusters.}
  \label{fig:projection}
\end{figure}

\noindent\textbf{Palmer Penguins --} We use the Palmer Penguins dataset as a proof-of-concept case to assess how well each of the three strategies reproduces simple cluster structures largely separable by well-documented~\cite{Marcus2023Palmer} biological attributes such as island, sex, and culmen dimensions. The five clusters under investigation are marked in Fig.~\ref{fig:projection}, with their dominant features summarized in Table~\ref{tab:penguins}. Besides compact Gentoo and Chinstrap groups, the dataset also includes more difficult cases: the blue bottom cluster, which separates into three Adelie subgroups by island, and the mixed middle cluster that combines individuals from several species.

Across strategies, the main discriminative features were generally recovered, while \textit{S1} most reliably reproduced the discriminative patterns across clusters. For Gentoo (Red top/right), the LLM reported larger body mass (mean 5{,}000--5{,}200,g) and flipper length (217--220,mm) together with lower culmen depth. For Chinstrap (Yellow left), higher culmen depth (18.4,mm) and shorter flippers (195,mm) were consistently captured. For the more difficult cases, only \textit{S1} reflected the even island split of the bottom cluster, aligning with Table~\ref{tab:penguins}. But like \textit{S2--3}, it did not identify the multi-species nature of the mixed cluster, although mentioning ``broad variability''.

\textit{S2} captured the main discriminative features of Gentoo and Chinstrap (e.g., large Gentoo body mass, deeper Chinstrap bills) but sometimes omitted secondary attributes like sex ratios. Reported ranges were less reliable and often expressed in vague terms such as ``extreme biometric values,'' and they did not consistently align with the ground-truth values. For the bottom cluster, it only mentioned variability, and for the mixed cluster, it gave rather vague labels such as ``mixed set,'' without reference to island subdivisions.

\begin{table}[ht]
\small
\centering
\caption{Manual, distribution-based comparison of various clusters.}
\vspace{-1mm}
\begin{tabular}{lcccccc}
\hline
\textbf{Attribute} & \textbf{Yellow(l.)} & \textbf{Red(t.)} & \textbf{Red(r.)} & \textbf{Blue(d.)} & \textbf{Mixed(m.)}\\
\hline
Island & Dream & Biscoe & Biscoe & Mixed & Dream \\
Culmen length & $\uparrow$ & $\uparrow$ & $-$ & $\downarrow$ & $-$ \\
Culmen depth & $\uparrow$ & $\uparrow$ & $-$ & $\downarrow$ & $-$ \\
Flipper length & $-$ & $\uparrow$ & $\uparrow$ & $\downarrow$ & $\downarrow$ \\
Body mass & $-$ & $\uparrow$ & $\uparrow$ & $-$ & $\downarrow$ \\
Sex & Male & Male & Female & Male & Female \\
\hline
\end{tabular}
\begin{flushleft}
\footnotesize
\vspace{-1mm}
\textit{Note:} Yellow = Chinstrap, Red = Gentoo, Blue = Adelie; $\uparrow$ = High, $-$ = Average, $\downarrow$ = Low.
\end{flushleft}
\label{tab:penguins}
\vspace{-6mm}
\end{table}

\textit{S3} produced slightly richer narratives with more contextual framing. For example, in the yellow cluster it described the non-selected set as having ``many smaller individuals,'' and in the mixed cluster it referred to ``morphological diversity.'' These phrasings align with the direction of the ground-truth contrasts but remain more qualitative and less precise than the statistical outputs. For the blue cluster, it noted variability but not the island-based subgroups.

\begin{figure*}[t]
  \centering
  \includegraphics[width=0.83\textwidth]{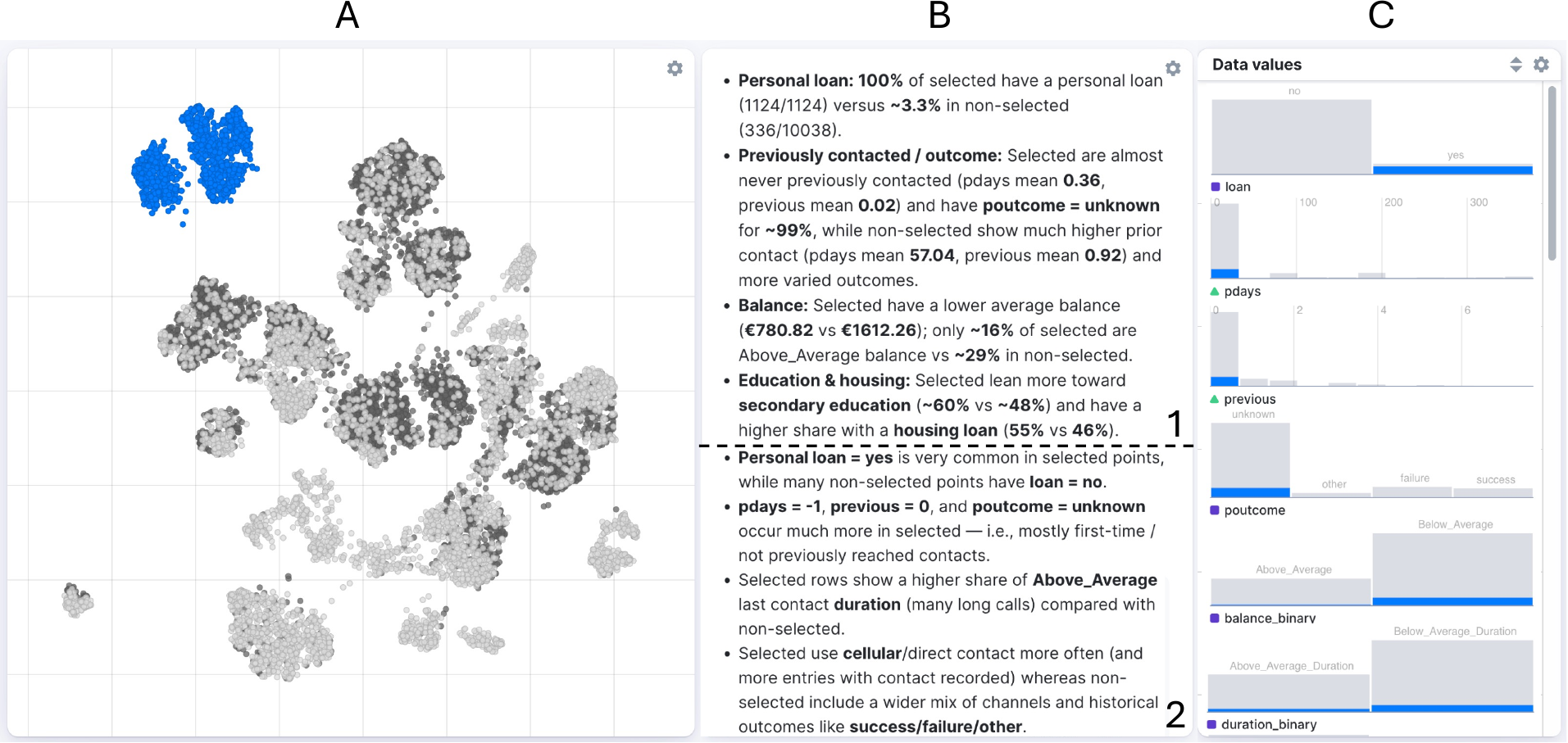}
  \vspace{-3mm}
  \caption{Further analysis of the \textit{bank marketing} dataset. (A) Projection of a distinct, manually selected cluster containing mixed class labels. (B.1) The LLM’s response using the statistics strategy is highly accurate. (B.2) The LLM’s response using the subsampling strategy is vague, limiting human understanding. (C) Data values are used as a validation method to assess the correctness of the LLM’s responses.}
  \label{fig:bank}\vspace{-4mm}
\end{figure*}

\noindent\textbf{Bank Marketing --} We also tested our approach on the larger Bank Marketing dataset. To verify robustness, we examined the ``yes'' label of the target attribute (\textit{deposit}) and confirmed that the LLM replicated the human analysis~\cite{Varun2023Bank}. However, \textit{S3} could not be applied due to token limits, presenting a systematic challenge in passing full datasets to LLMs. In both \textit{S1} and \textit{S2}, the LLM accurately identified above-average \textit{duration} and \textit{balance}, along with successful past campaigns (\textit{poutcome}), as strong indicators of subscription.

Beyond the shared findings, \textit{S1} provided more precise and comprehensive insights when numerical data were available, such as exact \textit{duration} (537s vs 223s) and \textit{balance} (€1{,}804 vs €1{,}280) values rather than using the binary attributes. Consistent with the human analysis, \textit{S1} repeatedly captured the strong link between not having a \textit{housing} loan and higher subscription rates (57\% vs 36.6\%), whereas \textit{S2} missed this pattern entirely. In contrast, \textit{S2} tended to describe patterns only qualitatively (e.g., more often,'' overwhelmingly''), limiting interpretability and comparability across analyses. Both strategies omitted a potentially relevant feature, \textit{age}, which was examined in the human analysis; however, \textit{S1} uncovered distinctive attributes absent from \textit{S2} and the human analysis, including tertiary-level \textit{education} (appearing only once) and predominance of ``cellular'' \textit{contact}.

Fig. \ref{fig:bank} presents a more exploratory analysis of this dataset, where we select a distinct cluster as shown in (A) and compare LLM outputs (B.1 vs B.2). Across three trials, both \textit{S1} and \textit{S2} produced generally accurate and consistent attributes, yet notable discrepancies remain: in (B.2), \textit{pdays} is incorrectly reported as --1 (actual mean 0.36, see (B.1)), and the model emphasizes \textit{duration} over the more discriminative \textit{balance} attribute (C). Notably, when numerical values are reported \textit{S1}, they are always consistent across trials (see supplementary material for details). In (B.1), \textit{S1} provides precise statements---e.g., correctly noting that \textit{100\%} of the selected points have a personal loan---whereas \textit{S2} uses vague qualifiers such as ``is very common'' or ``much more common,'' which reduce clarity and trustworthiness (B.2). Moreover, the \textit{balance} attribute, a distinctive feature with a clear separation in the distribution shown in (C) (second-to-last bar chart), is entirely missing from the \textit{S2} in one trial (see B.2), despite being reported with even more accurate values in B.1 (€780.82 vs €1612.26 in all trials).

\noindent\textbf{Food Nutrition --} In Fujiwara et al.’s work~\cite{Fujiwara2020Supporting}, the nutrient dataset is clustered and analyzed using ccPCA, where each feature’s relative contribution to the contrast between one cluster and the others is computed and visualized with heatmaps. For each cluster, their method highlights two or three features with the highest contributions, with particular emphasis on `calories' and `fat'. 

In our experiment, we generate LLM-based explanations for the same clusters. Rather than focusing primarily on a few features, the LLM provides more comprehensive explanations of how multiple features collectively contribute to distinguishing one cluster from the others, which aligns with the patterns observed in the feature distribution view. Furthermore, the LLM has the potential to categorize features and summarize results in an accessible way. For example, for one cluster, the LLM states that ``the selected subset differs from the non-selected group mainly by higher mineral and fat-related values...,'' where features such as calcium and sodium are automatically categorized as minerals, and fat and saturated fat are grouped as fat-related values.

\noindent\textbf{Customer Analysis --} We find that most salient differences between clusters are reliably captured, particularly in income, spending, and household composition, as in the human analysis~\cite{Karnika2021Customer}. For example, Cluster 1 households were consistently described as non-parents with higher income (75k vs 45k) and much higher spending (1,385 vs 388), while Cluster 2 was sharply identified as younger, low-income families with minimal spending (99 vs 788). Similarly, in Cluster 0, strong contrasts in purchase activity were highlighted (Wines 471 vs 227; NumStore 7.75 vs 4.89), and Cluster 3 was distinguished by very low campaign acceptance (4.6\% vs 17.6\%) together with reduced spending. These repeated patterns show that the separation between clusters is expressed in both demographic and behavioral terms, often with substantial numeric gaps.

Across all strategies, the differences lay less in which features were identified and more in how they were conveyed. \textit{S1} provided precise numeric contrasts that made the cluster boundaries explicit; for instance, ``Income $\approx$30k vs $\approx$60k'' for Cluster 2 and ``Response $\approx$4.6\% vs $\approx$17.6\%'' for Cluster 3. \textit{S2} tended to generalize these differences with vaguer ranges (``much less spending,'' ``lower income'') and occasionally added unsupported themes such as stronger promotion acceptance or different recency patterns. \textit{S3} leaned toward narrative framing, sometimes enhancing accessibility but at the expense of sharpness. For Cluster 1, for example, the description ``many high-income couples without children'' matched the demographic split but did not convey the exact magnitude of the income gap. Lastly, Cluster 0 was described as ``parents with teenagers and higher spending,'' which reflects the ground truth but omits the large purchase-level differences shown in the statistics.

\section{Discussion}   
\noindent\textbf{Experimental Results --} We find that \textit{S1} produces precise, consistent, and interpretable outputs by reporting verifiable measures such as means, standard deviations, and test values. This strategy yields the cleanest descriptions and makes discriminative features explicit, which makes it the most reliable approach for systematic analysis. The limitation is that the resulting text is more technical and less accessible to non-experts, but this does not reduce its analytical value.  
\textit{S2} does reasonably well but introduces ambiguity and risks omitting key discriminative features. Outputs often rely on vague language, for example, describing a feature as “being higher” rather than reporting absolute separations, and they are prone to overlooking important features.  
\textit{S3} generates richer narratives that can be more readable for non-technical users, capturing broader patterns such as extremes, ranges, or diversity. However, these outputs are often verbose and emphasize less relevant variation, and in 2 of the 4 datasets, \textit{S3} could not be applied due to token processing limits, which prevent a systematic one-to-one comparison.

\noindent\textbf{Reliability Considerations --} As with any LLM-based method, reliability is a central concern. The main risks are hallucinations, where the model introduces nonexistent features; overgeneralization, where weak signals are overstated as strong trends; and misleading fluency, where persuasive language masks factual inaccuracies. We address these issues by restricting prompts to observed attributes, avoiding speculative text and extrapolation beyond the selected cluster, and stating model limitations in the interface.


\noindent\textbf{Limitations and Future Work --} 
First, we did not explore combinations of the three strategies; future work could investigate hybrid approaches that merge the statistical precision of aggregated evidence with the richer narratives derived from the actual data. However, \textit{S3} was in most cases infeasible due to LLM token limits, and arguably providing the entire dataset to an LLM is not the most computationally- and energy-efficient approach. Instead, future work could explore methods that distill or abstract the dataset into concise (visual) representations---beyond simple statistics---that still capture its main patterns and relationships. Our evaluation was limited to four structured tabular datasets from specific domains; expanding to non-tabular, mixed-type, or more complex datasets would help assess generalizability. The number of sentences and bullet points in the explanations was intentionally kept short to match the needs of novice users, yet future work could explore adaptive settings that adjust detail based on user preference or task complexity. Finally, while our analysis compared outputs against ground-truth statistics, we did not conduct user studies; designing controlled experiments with novices and experts will allow us to measure the effect of each strategy on interpretation quality and confidence during decision making.

\section{Conclusion}
LangLasso provides an effective entry point for exploratory analysis by offering quick and accessible cluster interpretations. While not a definitive explanation method, it is well-suited for an early-stage analysis, teaching contexts, and novice users, where it can lower the entry barrier to data exploration. 

 \section*{Acknowledgements}
 This paper is a result of the discussion group work at ELLIIT Focus Period "Visualization-Empowered Human-in-the-Loop Artificial Intelligence" at Linköping University, Campus Norrköping, Sweden, Spring 2025, and financial support by REGATE (13N17427) and Austrian Science Fund (FWF) project 10.55776/COE12.



\bibliographystyle{abbrv-doi}
\bibliography{langlasso}
\end{document}